\documentclass{article}
\usepackage{spconf,amsmath,graphicx,hyperref}
\usepackage{amssymb}
\usepackage{bm}
\usepackage{xcolor}

\title{Distributed Multichannel Active Noise Control with Asynchronous Communication}
\name{Junwei Ji$^1$, Dongyuan Shi$^2$, Boxiang Wang$^1$, Ziyi Yang$^1$, Haowen Li$^1$, Woon-Seng Gan$^1$ \thanks{The code will be available at https://github.com/Ji-Junwei/ACDMCANC}}
\address{$^1$School of Electrical and Electronic Engineering, Nanyang Technological University, Singapore\\
         $^2$School of Marine Science and Technology, Northwestern Polytechnical University, Xi'an, China\\
         Email: JUNWEI002@e.ntu.edu.sg}
\begin{document}

\ninept
\maketitle
\begin{abstract}
Distributed multichannel active noise control (DMCANC) offers effective noise reduction across large spatial areas by distributing the computational load of centralized control to multiple low-cost nodes. Conventional DMCANC methods, however, typically assume synchronous communication and require frequent data exchange, resulting in high communication overhead. To enhance efficiency and adaptability, this work proposes an asynchronous communication strategy where each node executes a weight-constrained filtered-x LMS (WCFxLMS) algorithm and independently requests communication only when its local noise reduction performance degrades. Upon request, other nodes transmit the weight difference between their local control filter and the center point in WCFxLMS, which are then integrated to update both the control filter and the center point. This design enables nodes to operate asynchronously while preserving cooperative behavior. Simulation results demonstrate that the proposed asynchronous communication DMCANC (ACDMCANC) system maintains effective noise reduction with significantly reduced communication load, offering improved scalability for heterogeneous networks.

\end{abstract}
\section{Introduction}
\label{sec:intro}
Noise control has attracted increasing attention in recent years, as prolonged exposure to excessive noise poses risks to both physical and mental health. Among various approaches, active noise control (ANC) is particularly effective for attenuating low-frequency noise by generating an anti-noise signal with the same amplitude but opposite phase \cite{kuo2002active}. To cope with diverse noise types and dynamic acoustic environments, the filtered reference least mean squares (FxLMS) algorithm has become one of the most widely adopted solutions in ANC systems \cite{morgan2003analysis}. A wide range of FxLMS modifications have been developed to mitigate challenges encountered in real-world implementations \cite{lam2021ten,guo2024survey,su2025co,tsuji2025novel,zhang2025spherical}. More recently, advances in artificial intelligence (AI) have spurred the design of deep-learning-driven ANC algorithms \cite{zhang2021deep,xie2024cognitive,pike2025dynamic}, which provide promising avenues for achieving greater robustness and adaptability in complex acoustic conditions \cite{shen2025data,shi2024behind,luo2025frequency,wang2025transferable}.

Multichannel active noise control (MCANC) systems, which employ multiple secondary sources and error sensors, have attracted considerable attention for their ability to achieve global noise reduction over large spatial regions. A commonly adopted approach is the centralized strategy, in which a single processor handles all inputs and outputs while updating the control filters. Although effective, this method imposes a heavy computational burden, challenging the processor’s capacity and significantly increasing implementation costs \cite{Elliott1987MEANC}. In contrast, the decentralized strategy distributes the computation across multiple low-cost controllers, thereby reducing the load on any individual processor. However, this approach often suffers from instability issues arising from mutual acoustic crosstalk effect \cite{Elliott1994InteractionMANC}. To overcome these limitations, distributed control strategies have been increasingly explored, which aim to balance computational efficiency with system stability \cite{ferrer2015active,Song2016Diffusion}.

The distributed multichannel ANC (DMCANC) system comprises multiple ANC nodes, each running a single-channel ANC algorithm while exchanging key information to achieve global noise reduction. The diffusion FxLMS (DFxLMS) algorithm was developed to support such systems by employing topology-based combination rules for data fusion \cite{Chu2019DiffusionANC,Chu2020DiffusionANC}. However, DFxLMS essentially performs spatial smoothing \cite{Chu2021Combination}, which limits its effectiveness in handling asymmetric acoustic paths. To address this issue, the augmented DFxLMS (ADFxLMS) algorithm was proposed \cite{Li2023Distributed}, followed by a bidirectional communication strategy aimed at reducing communication overhead \cite{Li2023AugmentedDiffusion}. More recently, the auto shrink step size mixed gradient FxLMS (ASSS-MGDFxLMS) algorithm incorporated compensation filters to mitigate crosstalk and address communication latency \cite{Ji2025MGDFXLMS}. Nevertheless, most existing approaches assume data exchange at every sampling point, overlooking hardware limitations that prevent communication frequencies from matching ANC system sampling rates. In practice, communication issues significantly affect performance, revealing a gap between theoretical developments and real-world implementation \cite{li2024experimental}.

Therefore, we propose an asynchronous communication DMCANC (ACDMCANC) system. In this framework, each ANC node executes the weight-constrained FxLMS (WCFxLMS) algorithm \cite{Ji2025SBWCFXLMS} to mitigate instability arising from acoustic crosstalk during non-communication phases. Communication is triggered adaptively: a node requests information exchange only when its own local noise reduction performance ceases to improve. In response, the other nodes transmit the weight difference between their current control filter and the corresponding center point in the WCFxLMS. These differences are then combined through the mixed weight difference (MWD) operation, which updates both the local control filter and the center point. It is worth noting that each node makes communication decisions independently, based solely on its local performance. This autonomy leads to asynchronous communication, allowing nodes to continue running WCFxLMS seamlessly without interruption. As a result, the proposed ACDMCANC system effectively reduces communication overhead while enhancing adaptability to heterogeneous and time-varying network conditions.

The remainder of this paper is structured as follows: Section \ref{sec:MCANC} briefly reviews the DMCANC system. Section \ref{sec:ACDMCANC} describes the algorithms that is applied to ACDMCANC system. In Section \ref{sec:sim}, numerical simulations are conducted to demonstrate the effectiveness of the proposed method. Finally, conclusions are drawn in Section \ref{sec:concl}.

\begin{figure}[!t]
    \centering
    \includegraphics[width = 0.85\columnwidth]{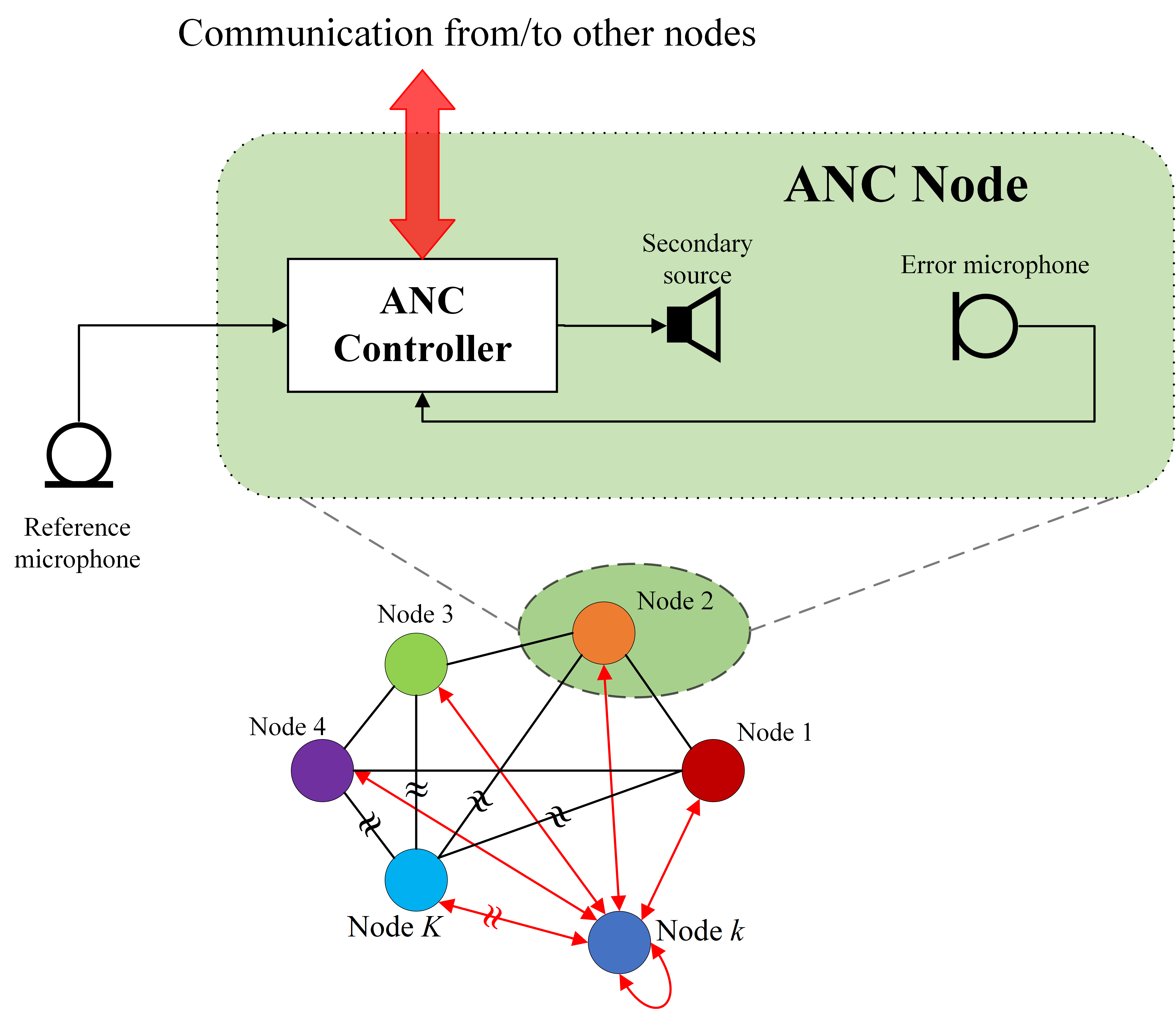}
    \caption{A DMCANC network, where each node consists of a secondary source, an error microphone, and an ANC controller for communication and processing, and shares one reference microphone.}
    \label{fig:ANCnode}
\end{figure}

\vspace{-0.5cm}
\section{Distributed MCANC System}\label{sec:MCANC}
{The MCANC system is widely employed to create large quiet zones. Among its control strategies, centralized control offers superior noise reduction performance but imposes a heavy computational burden on a single processor. In contrast, distributed control alleviates this burden by distributing computation across multiple nodes, though it requires inter-node communication to achieve effective noise reduction.}

Suppose the MCANC system consists of one reference sensor, $K$ secondary sources and $K$ error sensors. The reference sensor captures the reference signal $x(n)$ which is used to generate the $k$th control signal as
\begin{equation}\label{eq:controlsignal}
    y_k(n) = \mathbf{w}_k^\mathrm{T}(n)\mathbf{x}(n), \quad k = 1,2,...,K,
\end{equation}
where $\mathbf{w}_k(n)=[w_{k,0}(n) \, w_{k,1}(n) \, \cdots \, w_{k,L_w-1}(n)]^\mathrm{T}$ and $\mathbf{x}(n)=[x(n) \, x(n-1) \, \cdots \, x(n-L_w+1)]^\mathrm{T}$ denote the $k$th control filter and reference signal vectors with the length of $L_w$, and $n$ is the time index. The control signal $y_k(n)$ is reproduced by the $k$th secondary source and propagates through the secondary path $s_{mk}(n)$, generating the anti-noise that suppresses the disturbance signal $d_m(n)$. Accordingly, the $m$th error sensor measures the residual error as
\begin{equation}\label{eq:error}
    e_{m}(n) = d_{m}(n) - \sum_{k=1}^{K}{y}_{k}(n)*{s}_{mk}(n), \quad m = 1,2,...,K,
\end{equation}
where $*$ denotes the linear convolution, and ${s}_{mk}(n)$ refers to the impulse response of secondary path from the $k$th secondary source to the $m$th error sensor. Recall that the update equation of the conventional multiple error FxLMS (MEFxLMS) algorithm \cite{Elliott1987MEANC} is expressed as
\begin{equation}\label{eq:centralizedupdate}
    \mathbf{w}_k(n+1) = \mathbf{w}_k(n) + \mu \sum_{m=1}^K\mathbf{x}_{km}'(n)e_m(n),
\end{equation}
where $\mu$ denotes the step size, and $\mathbf{x}_{km}'(n)$ represents the filtered reference signal vector given by
\begin{equation}\label{eq:filteredx}
    \mathbf{x}_{km}'(n) = \hat{s}_{mk}(n)*\mathbf{x}(n),
\end{equation}
and $\hat{s}_{mk}(n)$ represents the estimate of the secondary path $s_{mk}(n)$.

The DMCANC system partitions this heavy computational load into multiple ANC nodes. Each node typically consists of a single controller, a secondary source, and an error sensor, as illustrated in Fig.~\ref{fig:ANCnode}. To derive the DMCANC algorithm, we first define the local gradient of the $k$th node as
\begin{equation}\label{eq:gradient}
    \boldsymbol{\nabla}_k(n) = \mu\mathbf{x}_{kk}'(n)e_k(n) =\mu[\mathbf{x}(n)*\hat{s}_{kk}(n)]\cdot e_k(n),
\end{equation}
According to \eqref{eq:filteredx} and \eqref{eq:gradient}, the centralized update equation in \eqref{eq:centralizedupdate} can be rewritten as
\begin{equation}\label{eq:cupdate}
    \mathbf{w}_k(n+1) = \mathbf{w}_k(n) + \boldsymbol{\nabla}_k(n) + \sum_{m=1,m\neq k}^K[\mathbf{x}(n)*\hat{s}_{mk}(n)]e_m(n).
\end{equation}
By comparing the last term in \eqref{eq:cupdate} with the definition of the local gradient in \eqref{eq:gradient}, it can be seen that the difference lies in the filtering process. Specifically, the local gradient is computed using the node’s self-secondary path $\hat{s}_{mm}(n)$, whereas the centralized update involves the cross-secondary paths $\hat{s}_{mk}(n)$ ($m \neq k$).

To reconcile this mismatch, compensation filters $c_{mk}(n)$ are introduced to approximate the difference between self and cross-secondary paths \cite{Ji2025MGDFXLMS}, expressed as
\begin{equation}\label{eq:compensate}
 \hat{s}_{mk}(n) = \hat{s}_{mm}(n) * c_{mk}(n), \; (m\neq k).
\end{equation}
In practice, there are $K(K-1)$ compensation filters in total, which can be pre-estimated through offline training before ANC operation. Substituting \eqref{eq:compensate} into \eqref{eq:cupdate}, and recalling the definition of the local gradient in \eqref{eq:gradient}, the update equation becomes
\begin{equation}\label{eq:mgdfxlms}
            \mathbf{w}_k(n+1) = \mathbf{w}_k(n) + {\boldsymbol{\nabla}_k(n)}+{\sum_{m=1,m\neq k}^K \boldsymbol{\nabla}_m(n)*c_{mk}(n)},   
\end{equation}
where $\boldsymbol{\nabla}_m(n)$ denotes the local gradient of node $m$, which is exchanged among nodes through inter-node communication. This formulation constitutes the mixed gradient FxLMS (MGDFxLMS) algorithm \cite{Ji2025MGDFXLMS}, which extends the conventional centralized MEFxLMS to a distributed setting.

\vspace{-0.5cm}
\section{Asynchronous communication DMCANC}\label{sec:ACDMCANC}
Conventional DMCANC systems require each ANC node to communicate at every sampling instant, which is impractical in real-world applications due to excessive communication demands. To address this limitation, we propose an asynchronous communication DMCANC (ACDMCANC) system, where each ANC node communicates independently based on its local condition. This strategy not only reduces the overall communication overhead but also enhances the system’s flexibility and scalability.
\begin{figure}[!t]
    \centering
    \includegraphics[width = 0.85\columnwidth]{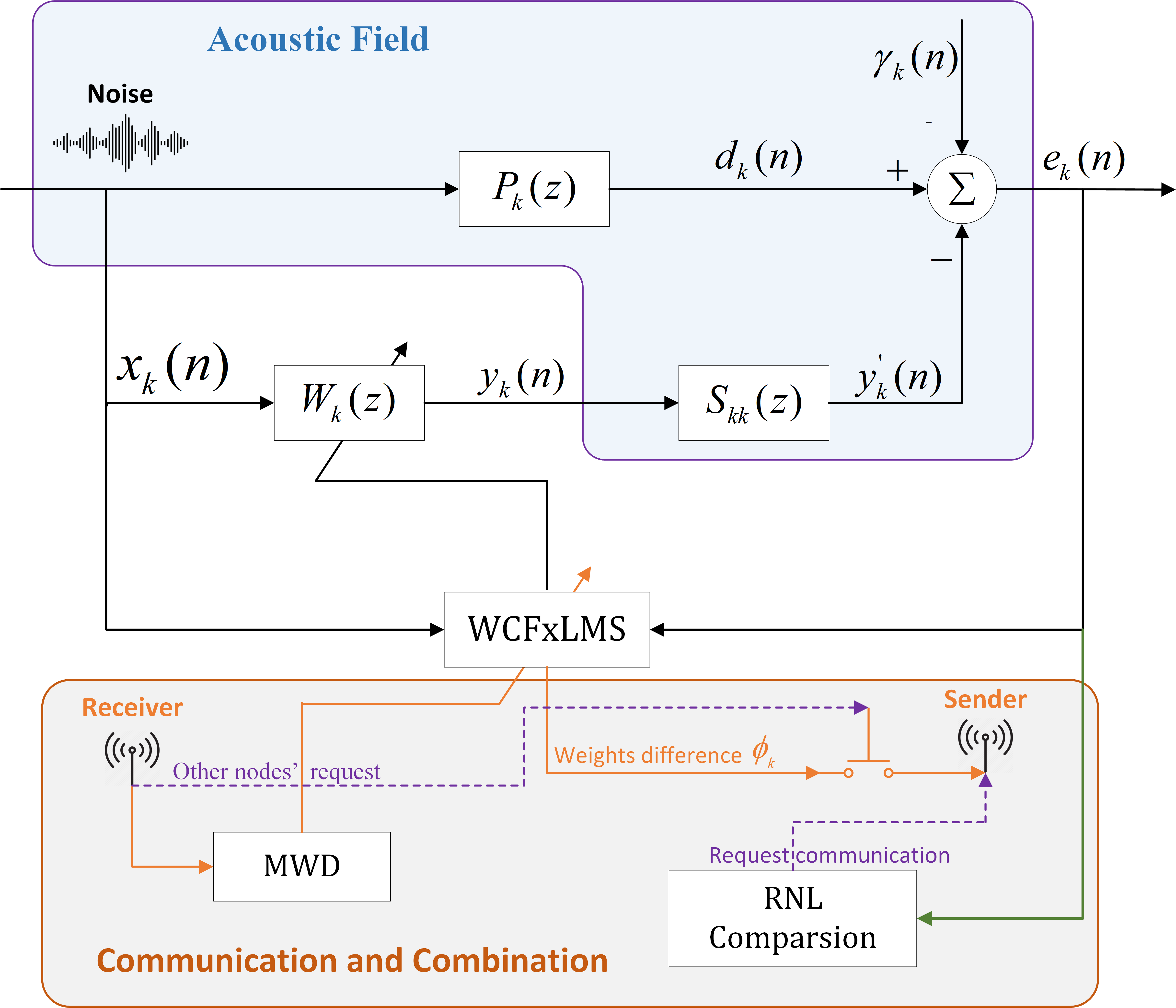}
    \caption{The block diagram of the $k$th node for the proposed ACDMCANC system, where $P_k(z)$ represents the primary path. $\gamma_k(n)$ is the interference caused by the other nodes, resulting in crosstalk between nodes \cite{Ji2025SBWCFXLMS}.}
    \label{fig:ACDMCANC}
\end{figure}

\subsection{Weight-constrained FxLMS}
Due to discontinuous communication, the update range of the control filter should be constrained to prevent instability caused by crosstalk among nodes during non-communication phases. To achieve this, the cost function of each node is modified as
\begin{equation}\label{eq:costfunction}
    J_k = \mathbb{E}[{e}^2_{k}(n)] + \alpha||\widetilde{\mathbf{w}}_k-\mathbf{w}_k(n)||^2,
\end{equation}
where $\alpha$ denotes the penalty factor, $|\cdot|$ represents the 2-norm operator, and $\widetilde{\mathbf{w}}_k$ is the constrained center point. Following the corresponding update rule, the control filter is updated as
\begin{equation}\label{eq:WCFxLMS}
    \mathbf{w}_k(n+1) = \mathbf{w}_k(n) + \mu \mathbf{x}_{kk}'(n)e_k(n) + \mu \alpha(\widetilde{\mathbf{w}}_k-\mathbf{w}_k(n)).
\end{equation}
Equation \eqref{eq:WCFxLMS} defines the weight-constrained FxLMS (WCFxLMS) algorithm \cite{Ji2025SBWCFXLMS}, where the last term serves as a penalty that restricts the maximum deviation of the adaptive filter from its center point. Consequently, even in the presence of strong crosstalk impulses, the control filter can only deviate within a bounded range, thereby preventing divergence.

\subsection{Asynchronous communication strategy}
Since the WCFxLMS algorithm limits the update range of the control filters, it effectively prevents divergence caused by inter-node crosstalk during the non-communication phase. However, this constraint inevitably reduces the maximum achievable noise reduction performance. To overcome this limitation, a suitable communication strategy is required so that global information can be periodically integrated, thereby enhancing overall system performance. To reduce the communication burden, the proposed system adopts an asynchronous communication strategy, in which each node independently determines whether to request communication. The decision is based on the residual noise level (RNL), defined for the $k$th node as
\begin{equation}\label{eq:nrl}
    \eta_k(n) = 10\log_{10}\left[{e^2_k(n)}\right].
\end{equation}
Due to the slow convergence of the LMS algorithm, an instantaneous evaluation may not adequately reflect performance. Hence, an updating period $T$ is introduced, over which the average RNL (ANRL) is computed as
\begin{equation}\label{eq:anrl}
    \bar{\eta}_k = \frac{1}{Tf}\sum_{t=n-Tf}^{n} \eta_k(t),
\end{equation}
where $f$ denotes the sampling frequency. If the current ARNL is worse than its previously recorded value, the node initiates a communication request.

Upon receiving the request, the other nodes transmit their weight difference, defined as the deviation between the updated local control filter and the center point:
\begin{equation}\label{eq:WD}
    \boldsymbol{\phi_k}(n) = \mathbf{w}_k(n+1) - \widetilde{\mathbf{w}}_k.
\end{equation}
This weight difference $\boldsymbol{\phi}_k(n)$ implicitly accumulates the effect of multiple local gradient updates and thus acts as a compact representation of the local adaptation history. Inspired by the MGDFxLMS algorithm expressed in \eqref{eq:mgdfxlms}, we extend the combination rule by replacing the local gradient $\boldsymbol{\nabla}_k(n)$ with $\boldsymbol{\phi}_k(n)$, leading to the mixed weight difference (MWD) update:
\begin{equation}\label{eq:MWD}
    \widetilde{\mathbf{w}}_{k,new} =   \widetilde{\mathbf{w}}_k + \boldsymbol{\phi_k}(n) + \sum_{m=1,m\neq k}^{K} \phi_m(n)*c_{mk}(n).
\end{equation}
This new control filter $\widetilde{\mathbf{w}}_{k,new}$ will then update both local control filter $\mathbf{w}_k(n)$ and center point $\widetilde{\mathbf{w}}_k$ in the WCFxLMS for further iteration.

The block diagram of the proposed ACDMCANC system is shown in Fig.~\ref{fig:ACDMCANC}. During non-communication phases, each node updates its control filter using the WCFxLMS algorithm, thereby maintaining a certain degree of local noise reduction and preventing instability. When communication is triggered, the exchanged weight differences from other nodes are fused through the MWD operation, enabling the system to improve overall performance. By allowing nodes to communicate only when necessary and ensuring that non-communicating nodes remain adaptive, the proposed ACDMCANC system significantly reduces communication overhead while improving adaptability to heterogeneous network conditions. This design ensures that the network can balance stability, communication efficiency, and global noise reduction performance. {For completeness, a synchronous DMCANC (SCDMCANC) variant is also considered in this work. It employs the same WCFxLMS adaptation and MWD combination rules as the proposed ACDMCANC, differing only in the communication strategy, where all nodes communicate simultaneously once any node triggers a communication event.}

\vspace{-0.5cm}
\section{Numerical simulations}\label{sec:sim}
In this section, numerical simulations are conducted to evaluate the performance of ACDMCANC system. The primary and secondary acoustic paths are measured in a real noise chamber equipped with an ANC window. The system configuration follows the setup in \cite{Ji2025MGDFXLMS}, consisting of six ANC nodes. The lengths of the secondary paths, compensation filters, and control filters are set to $256$, $33$, and $512$, respectively. The sampling frequency is $16{,}000$ Hz. The center point $\widetilde{\mathbf{w}}_k$ is initialized to zero, corresponding to the leaky FxLMS algorithm \cite{kuo2002active}, which allows the penalty factor to be selected by the conventional rule.

The proposed ACDMCANC system is compared against three representative approaches: (i) the centralized MEFxLMS algorithm \cite{Elliott1987MEANC}, (ii) the MGDFxLMS algorithm \cite{Ji2025MGDFXLMS}, and (iii) the SCDMCANC system. For simplicity, we assume that there is no transmission delay between nodes. To quantitatively assess noise reduction performance, the average normalized squared error (ANSE) across all ANC nodes is defined as
\begin{equation}\label{eq:anse}
    \text{ANSE}(n) = \frac{1}{K}\sum_{k=1}^{K} 10\log_{10}\bigg\{\frac{\mathbb{E}[e^2_k(n)]}{\mathbb{E}[d^2_k(n)]}\bigg\},
\end{equation}
where the expectation is calculated by averaging the signal over $5,000$ samples.

\subsection{Broadband noise reduction performance}
\begin{figure}[!t]
    \centering
    \includegraphics[width = 0.9\columnwidth]{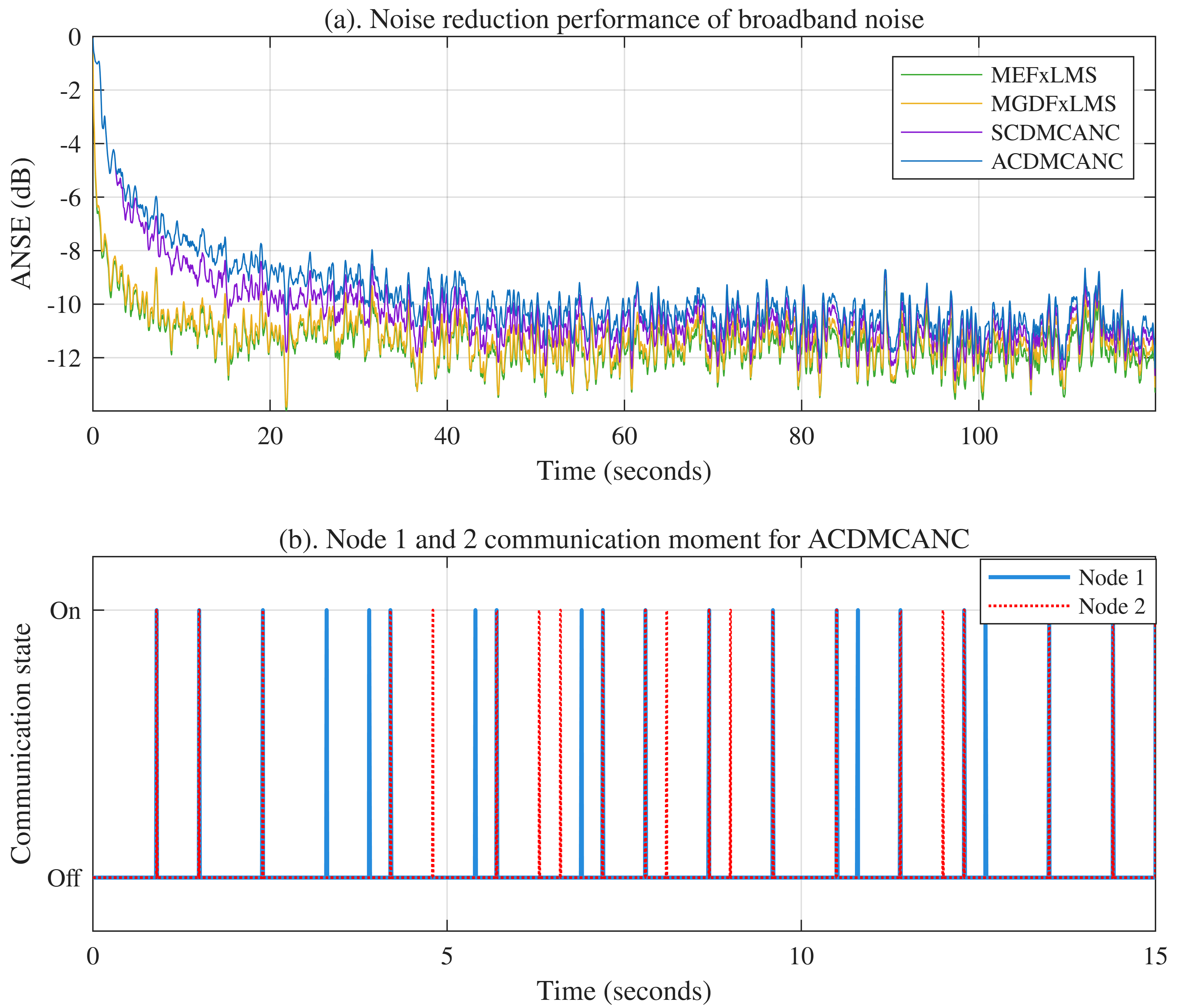}
    \caption{Noise reduction performance for broadband noise using different algorithms: (a). ANSE comparison of various MCANC algorithms. (b) illustrate the communication instances of Node 1 and Node 2 in the ACDMCANC system, respectively, where “Off” indicates no communication and “On” denotes a communication request.}
    \label{fig:case1}
\end{figure}
In this case study, the primary noise is broadband noise ranging from $100$ to $1,000$Hz to evaluate the proposed method. The step size for all algorithms is fixed at $1\times10^{-6}$, and the penalty factor is set to $800$ for all nodes. The time interval for calculating the RNL is $T=0.3$s. As shown in Fig.~\ref{fig:case1}(a), the MGDFxLMS achieves nearly the same noise reduction performance as the centralized MEFxLMS, since it is directly derived from the centralized algorithm with modifications suitable for distributed control. The SCDMCANC also attains comparable performance to both MGDFxLMS and MEFxLMS. However, because RNL must be periodically checked to determine whether communication is required, its communication frequency is reduced. Unlike MGDFxLMS, which communicates at every sampling point, this results in slower convergence. The proposed ACDMCANC achieves satisfactory noise reduction performance, although slightly inferior to the other methods. This limitation arises because, under asynchronous communication, once a node communicates and updates its control filter and constrained center point, subsequent requests from other nodes may transmit weight differences that miss certain accumulated historical gradient information. Nevertheless, the ACDMCANC still ensures effective global noise reduction while enabling independent communication for each node.

 Figures \ref{fig:case1}(b) show the communication records of Nodes 1 and 2 in ACDMCANC during the first $15$ s. The timing differences between the two nodes highlight the asynchronous nature of the system. These results confirm that the proposed WCFxLMS and MWD methods can operate effectively in heterogeneous networks, reducing communication frequency while maintaining flexibility and applicability across different network conditions.

\vspace{-0.5cm}
\subsection{Real recorded noise reduction performance}

\begin{figure}[!t]
    \centering
    \includegraphics[width = 0.9\columnwidth]{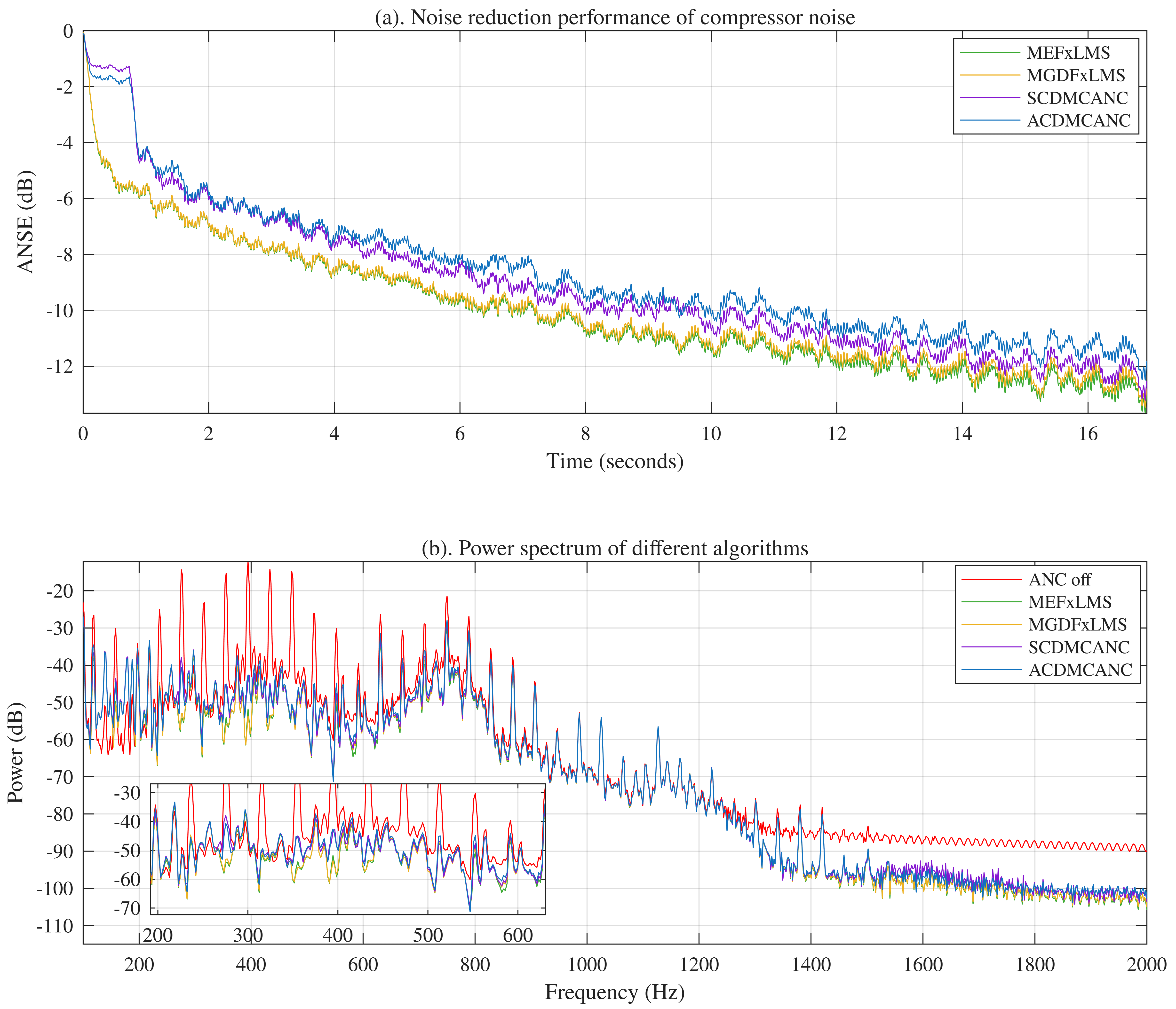}
    \caption{Noise reduction performance for real recorded compressor noise using different algorithms: (a). ANSE comparison of various MCANC algorithms; (b) Power spectrum comparison of various MCANC algorithms}
    \label{fig:case2}
\end{figure}

In this case, a real recorded compressor noise is used as the primary noise source. The step size for both SCDMCANC and ACDMCANC is set to $5\times10^{-6}$, while the penalty factor is adjusted to $400$. As shown in Fig.~\ref{fig:case2}, the MGDFxLMS achieves nearly identical noise reduction performance to the centralized MEFxLMS at the cost of high communication burden, which limits its feasibility in real-time applications. In contrast, SCDMCANC and ACDMCANC reduce communication overhead by allowing nodes to communicate adaptively, though at the expense of slower convergence and slightly lower final performance. Nevertheless, the incorporation of WCFxLMS and the MWD method ensures that both synchronous (SCDMCANC) and asynchronous (ACDMCANC) communication strategies can be flexibly adopted. In particular, asynchronous communication is advantageous in heterogeneous network environments, enhancing the universality and robustness of the DMCANC framework.

\vspace{-0.5cm}
\section{Conclusion}\label{sec:concl}

This paper introduced an asynchronous communication strategy for DMCANC (ACDMCANC), aiming to reduce the excessive communication burden inherent in conventional distributed systems. By integrating the weight-constrained FxLMS (WCFxLMS) algorithm with a mixed weight difference (MWD) combination rule, each node can maintain stability during non-communication phases while flexibly incorporating global information when communication is triggered. Simulation results demonstrate that the proposed method offers a promising balance between performance, communication efficiency, and robustness. Moreover, asynchronous communication provides greater adaptability to heterogeneous networks, which is essential for real-world deployment.


\bibliographystyle{IEEEbib}
\bibliography{refs}

\end{document}